\documentclass[a4paper,fleqn,usenatbib]{mnras}

\usepackage[T1]{fontenc}
\usepackage{ae,aecompl}

\usepackage{graphicx}	
\usepackage{amsmath}	
\usepackage{amssymb}	

\def\msun{{\rm\,M_\odot}}

\def\gtsima{$\; \buildrel > \over \sim \;$}
\def\simgt{\lower.5ex\hbox{\gtsima}}

\def\msun{\hbox{M$_\odot$}}
\title[The LMC star cluster age gap]{A genuine Large Magellanic Cloud age gap star cluster}

\author[Andr\'es E. Piatti]{
Andr\'es E. Piatti$^{1,2}$\thanks{E-mail: andres.piatti@unc.edu.ar}\\
$^{1}$Instituto Interdisciplinario de Ciencias B\'asicas (ICB), CONICET-UNCUYO, Padre J. Contreras 1300, M5502JMA, Mendoza, Argentina\\
$^{2}$Consejo Nacional de Investigaciones Cient\'{\i}ficas y T\'ecnicas, Godoy Cruz 2290, C1425FQB,  Buenos Aires, Argentina}

\date{Accepted XXX. Received YYY; in original form ZZZ}

\pubyear{2021}

\begin{document}
\label{firstpage}
\pagerange{\pageref{firstpage}--\pageref{lastpage}}
\maketitle

\begin{abstract}
We confirm the existence of a second Large Magellanic Cloud (LMC) star cluster, KMHK~1592,  
with an age that falls in the middle of the so-called LMC star cluster age gap, a long
period of time ($\sim$ 4 - 11 Gyr) where no star cluster had been uncovered, except ESO\,121-SC\,03.
The age  (8.0$\pm$0.5 Gyr) and the metallicity  ([Fe/H]=-1.0$\pm$0.2 dex) of KMHK~1592 were derived from the fit of theoretical isochrones to the
intrinsic star cluster colour-magnitude diagram sequences, which were unveiled using a robust 
star-by-star membership probability procedure. Because of the relative low brightness of the star 
cluster,  deep GEMINI GMOS images were used.  We discuss the pros and cons of three glimpsed 
scenarios that could explain the presence of both LMC age gap star clusters in the outskirts of 
the LMC, namely: in-situ star cluster formation, capture from the Small Magellanic Cloud, or 
accretion of a small dwarft galaxy. 
\end{abstract} 




\section{Introduction}

The absence of star clusters with ages between $\sim$ 4 and 11 Gyr in the
Large Magellanic Cloud (LMC), the sole exception is ESO\,121-SC\,03
\citep{mateoetal1986}, was noticed by \citet{oetal91}. They also found 
that the age gap correlates with a star cluster metallicity gap, in the sense that star clusters 
younger than 3 Gyr are much more metal-rich than the ancient LMC globular 
clusters. Although different observational campaigns have searched for unknown 
old LMC star clusters, they have confirmed previous indications that star clusters were 
not formed during the age gap \citep[][]{dc1991,getal97}.

The upper age limit of the LMC star cluster age gap is given by the youngest ages of the 15 
known LMC globular clusters \citep[$\sim$ 12 Gyr][]{betal08}.
The lower age limit, however, has been changed as more intermediate-age star clusters
were studied in detail. For instance, \citet{s98} found that NGC\,2121, 2155 and
SL\,663 are $\sim$ 4 Gyr old star clusters, while \citet{richetal2001}
 re-estimated their ages to be 0.8 Gyr younger. Age estimates of poorly studied or 
unstudied star clusters were derived during the last decade, and the oldest ones turned 
out to be $\sim$ 2.5-3.0 Gyr old \citep[see, e.g.][]{pg13}

The LMC star cluster age distribution was modeled by \citet{bekkietal2004}, who
proposed that the LMC was formed at a distance from the Milky Way that did not 
allow its tidal forces to trigger star cluster formation efficiently. The star cluster
formation resumed in the LMC at its first encounter with the Small Magellanic 
Cloud (SMC) $\sim$ 2-3 Gyr ago. Such a star cluster formation history
was not that of the SMC, which would have been formed
as a lower mass galaxy closer to the Milky Way, and thus more continuously
influenced by its gravitational field. Nevertheless, both Magellanic Clouds have
had a series of close interactions between them and with the Milky Way since then,
that  explain their abrupt observed chemical enrichment history and increase 
of the star cluster formation rates \citep{perrenetal2017}. For the sake of the 
reader we refer to some recent studies dealing with the LMC formation and interaction with the SMC, namely: \citet{baetal13,zivicketal2019,williamsetal2021,mazzietal2021,cullinaneetal2022}, among others.

\citet{petal14b} used a $K_s$ vs $Y-K_s$ colour-magnitude diagram (CMD) to 
estimate for the first time the age of KMHK~1592, a low surface brightness LMC star cluster 
located in the LMC outer disc (RA = 90.375$\degr$; Dec = -66.987$\degr$).
Although the CMD does not reach the star cluster Main Sequence turnoff, 
\citet{petal14b} estimated an age of 6.3 Gyr, which places KMHK~1592 in the 
middle of the age gap. The star cluster could be even older, for instance, of the age of the 
LMC globular clusters  ($\sim$ 12-13 Gyr).

Because of the discovery of only one LMC age gap star cluster
would be worthy by itself, this astonishing new age gap star cluster
candidate deserves our attention. 
Furthermore, \citet{piatti2021d} analysed
17 previously unidentified star cluster candidates with estimated ages $\ga$ 4 Gyr
\citep{gattoetal2020} and favoured from the analysis of their CMDs and spatial distribution maps
the existence of LMC change grouping stars, although a definitive assessment on 
them will be possible from further deeper photometry. 
As far as we are aware, KMHK\,1592 has not
been observed by any of the ongoing or recent LMC surveys 
(e.g., DES \citep{flaugheretal2015}, SMASH \citep{nideveretal2017a}, STEP \citep{retal14},
VISCACHA \citep{maiaetal2019}). Precisely, the aim of this 
Letter is to report new observations of KMHK~1592, from which we confirm it as 
the second genuine LMC age gap star cluster. In Section~2 we describe the observations
and several data processing, while
Section~3 deals with the analysis of the data obtained and the discussion of the resulting
star cluster age.

\begin{figure}
\includegraphics[width=\columnwidth]{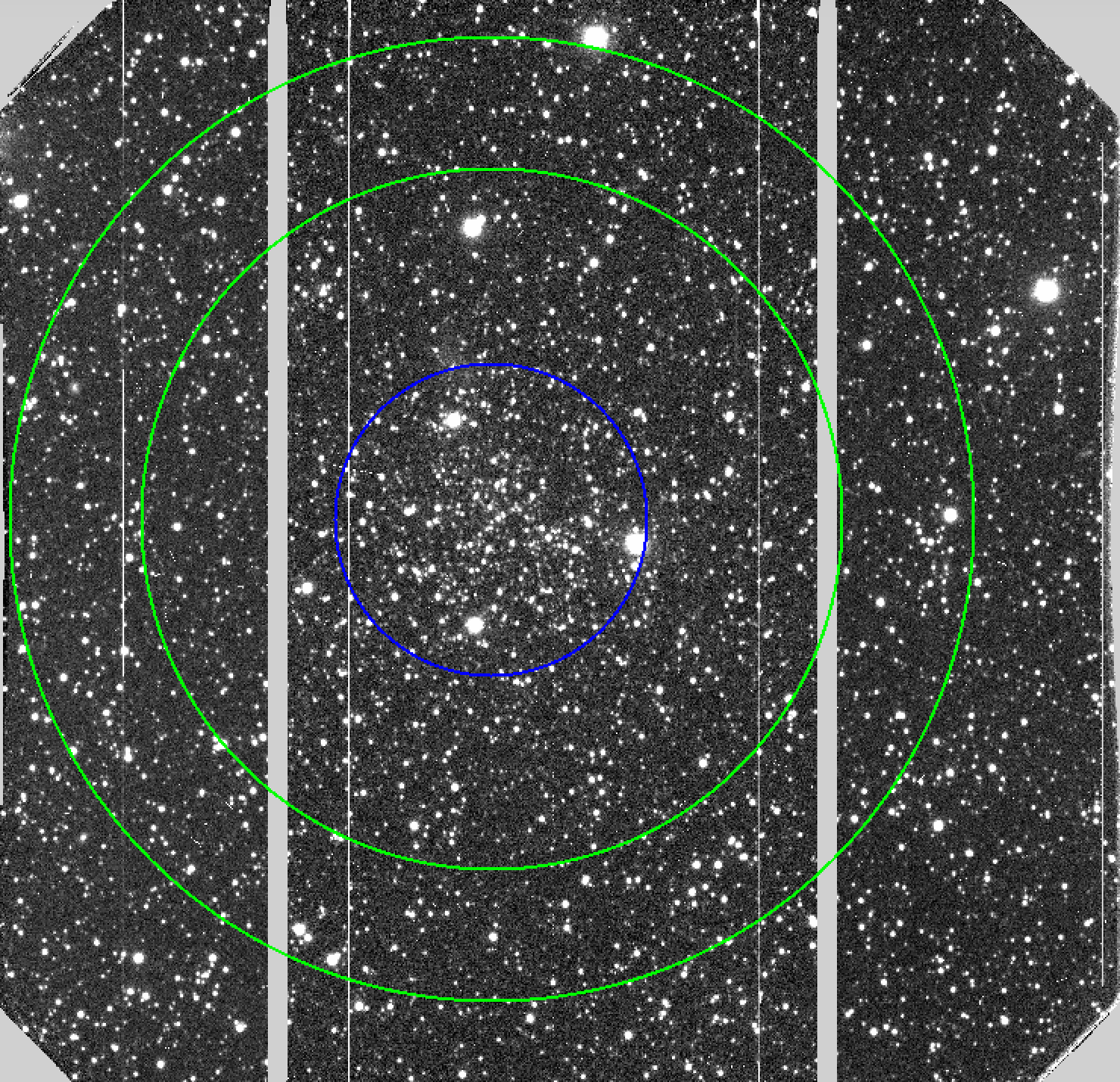}
\caption{GMOS $g$ image centred on KMHK~1592. Green circles delimite the
cluster circle and the external annulus used as reference during the CMD cleaning of
field stars. The blue circle represents the cluster's radius \citep{petal14b}. North is up
and east is to the left.}
\label{fig1}
\end{figure}

\section{Data collection and processing}

We carried out observations  of KMHK~1592 with the Gemini South telescope 
and the GMOS-S instrument  (3$\times$1 mosaic of 2K$\times$4K EEV CCDs)
through $g$ and $i$ filters on the night of November 25, 2021, under
program GS-2021B-FT-108. We obtained 3$\times$150 sec images per filter in 
excellent seeing (0.62$\arcsec$ to 0.93$\arcsec$ FWHM) and photometric 
conditions, at a mean airmass of 1.25. In order to transform the instrumental
magnitudes into the standard system, we obtained 3$\times$30 sec exposures
in $g$ and $i$, respectively, of NGC~2155, an LMC intermediate-age star cluster 
located $\sim$1.5$\degr$ from KMHK~1592 with standard $g,i$ photometry obtained 
by \citet{petal14}. The observations were performed
just after finishing those for KMHK~1592. The data reduction followed the procedures 
documented in the Gemini Observatory  webpage\footnote{http://www.gemini.edu}
and utilized the {\sc gemini/gmos} package in Gemini {\sc iraf}. We performed overscan, 
trimming, bias subtraction, flattened all data images, etc., once the
calibration frames (bias and flats) were properly combined.

We derived the stellar photometry of KMHK~1592 and NGC~2155 using the star-finding 
and point-spread-function 
(PSF) fitting routines in the {\sc daophot/allstar} suite of programs \citep{setal90}. For each 
frame, we obtained a quadratically varying  PSF by fitting $\sim$ 100 stars, once we 
eliminated the neighbours using a preliminary PSF derived from the brightest, least 
contaminated $\sim$ 40 stars. Both groups of PSF stars were interactively selected. We
then used the {\sc allstar} program to apply the resulting PSF to the identified stellar 
objects and to create a subtracted image which was used to find and measure magnitudes 
of additional fainter stars. This procedure was repeated three times for each frame. 
We combined all the independent $g,i$ instrumental magnitudes using the stand-alone 
{\sc daomatch} and {\sc daomaster} programs\footnote{Program kindly provided by P.B. Stetson}. 
As a result, we produced three data sets per star cluster containing
the $x$ and $y$ coordinates for each star, and the instrumental $g$ and $i$ magnitudes
($\tilde{g}$,$\tilde{i}$) with their
respective errors. We cross-matched the instrumental and standard photometries
of NGC~2155 and found 2671 stars in common, from which we obtained the following 
transformation equations:

\begin{equation}
g = (0.998\pm0.004)\times \tilde{g} + 6.535\pm0.007,  rms=0.031
\end{equation}

\begin{equation}
i = (1.025\pm0.003)\times \tilde{i} + 6.522\pm0.005, rms=0.028
\end{equation}

\noindent for 15.0 $<$ $g$ (mag) $<$ 25.0, where right and left terms
refer to instrumental and standard magnitudes, respectively, indicating excellent 
photometric quality. We finally  transformed instrumental magnitudes of KMHK~1592 
into standards ones from eqs. (1) and (2), and derived robust photometric uncertainties
from the average of the three independent photometric data sets.

A key tool to unveil the actual age of KMHK~1592 is the cluster CMD cleaned of
field star contamination. We applied a decontamination procedure
based on that devised by \citet{pb12}, which properly reproduces the composite
observed field star population, and assigns membership probability to each star.
In order to clean the cluster CMD, 
we need to compare it with that of a reference star field, and then to properly eliminate from
the former a number of stars equal to that found in the latter, bearing in mind that 
the magnitudes and colours of the eliminated stars in the cluster CMD must reproduce 
the respective magnitude and colour distributions in the 
reference star field. For that purpose, we traced an annulus in the GMOS field of view 
(see Fig.~\ref{fig1}) with an inner radius $\ga$ 2 times the KHMK~1592's
radius \citep[0.8$\arcmin$,][]{petal14b}, and an area equal to the inner circle 
($r$=1.8$\arcmin$). The
outer annulus was used as the reference star field, while the inner circle served as the
cluster circle to be cleaned.

\begin{figure}
\includegraphics[width=\columnwidth]{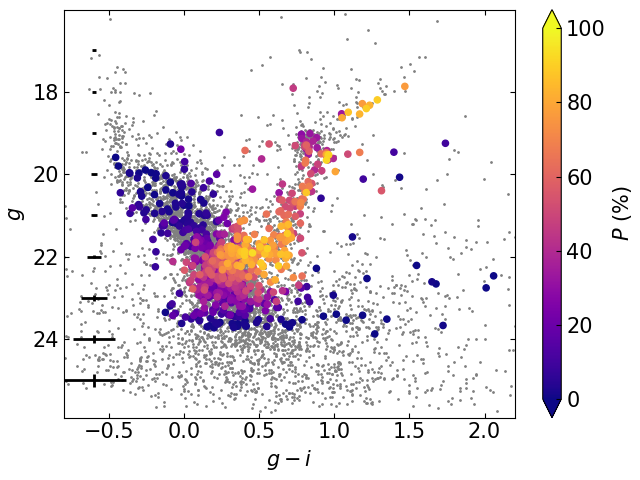}
\caption{CMD for all the measured stars represented with gray dots. Large
circles are stars located inside the cluster's radius \citep{petal14b}, coloured
according to their membership probability ($P$). Error bars are indicated
to the left.}
\label{fig2}
\end{figure}

The methodology to select stars to subtract from the cluster CMD
consists in defining boxes centred on the magnitude and colour of each star of the
reference star field; then to superimpose them on the star cluster CMD,
 and finally to choose one star per box to subtract. In order to guarantee  that a star 
is within the box boundary, we considered
boxes with size of ($\Delta$$g$,$\Delta$$(g-i)$) = (0.50 mag, 0.25 mag).
 In the case that more than one star 
is located inside a box, the closest one to its centre is subtracted. 
During the choice of the subtracted stars we took into account
their magnitude and colour errors by allowing them to have a thousand different
values of magnitude and colour within an interval of 
$\pm$1$\sigma$, where $\sigma$ represents the errors in their magnitude and colour, 
respectively. We also imposed the condition
that the spatial positions of the subtracted stars were chosen randomly. 
In practice, for each reference field star we randomly selected a
position in the cluster circle and searched for a star to subtract within a box of
0.2$\arcmin$ a side.
We iterated this loop up to 1000 times, if no star was
found in the selected spatial box. The outcome of the cleaning procedure
is a cluster CMD that likely contains only cluster  members. We executed 1000  times the decontamination procedure described above,
so that we obtained 1000 different cleaned CMDs. From them, we 
defined a membership probability $P$ 
($\%$) as the ratio $N$/10, where $N$ (between 0 and 1000) is the number 
of times a star was found among the 1000 different  cleaned CMDs. 

\begin{figure}
\includegraphics[width=\columnwidth]{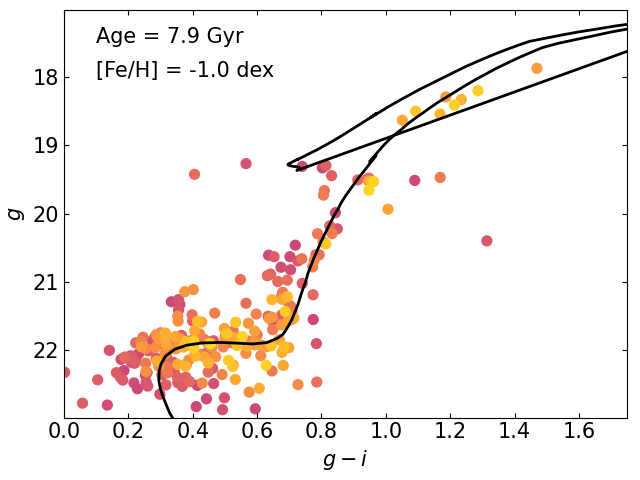}
\includegraphics[width=\columnwidth]{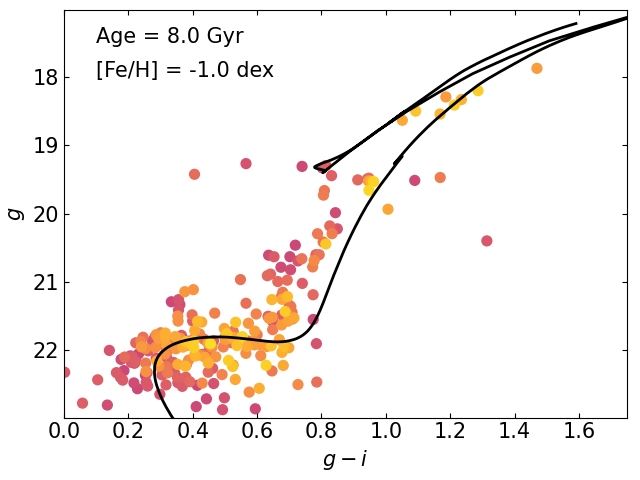}
\caption{CMD for stars with membership probability $P$ $>$ 50$\%$). The
isochrones by \citet{betal12} (top panel) and \citet{pietrinfernietal2004} (bottom),
 which best reproduce the cluster's sequences, are superimposed, respectively.}
\label{fig3}
\end{figure}

\begin{figure}
\includegraphics[width=\columnwidth]{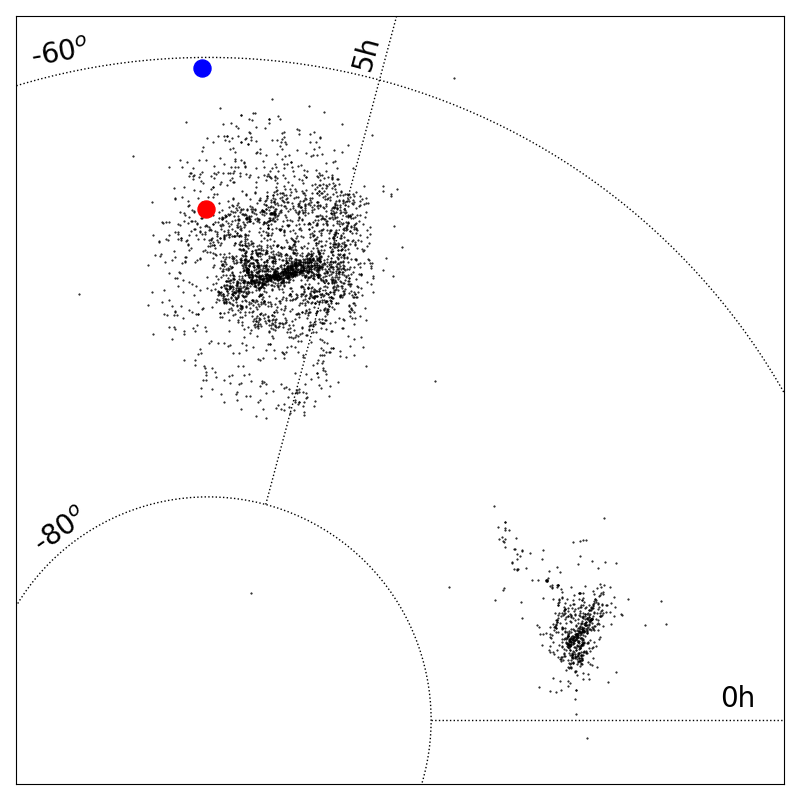}
\caption{Equal-area Hammer projection of the Magellanic Cloud star cluster
population in Equatorial coordinates \citep[small black dots][]{betal08}.
ESO\,121-SC\,03 and KMHK~1592 are represented by large blue and
red filled circles, respectively.}
\label{fig4}
\end{figure}

\section{Analysis and discussion}

Fig.~\ref{fig2} shows the observed CMD of KMHK~1592, where stars located
within the cluster radius are coloured according to their membership probability.
As can be seen, a red giant branch, a red clump,
a populous subgiant branch and a Main Sequence turnoff are clearly highlighted
as the most probable star cluster sequences. It is readily visible that the cluster
red clump nearly superimposes that of the composite star field population, which
implies that KMHK~1592 belongs to the LMC and is nearly at the mean LMC
distance \citep[49.9 kpc][]{dgetal14}. Likewise, \citet{petal14b} estimated a low cluster
reddening of $E(B-V)$ = 0.042$\pm$0.010 mag from the Magellanic Cloud extinction 
values based on red clump stars photometry provided by the OGLE collaboration 
\citep{u03} as described in \citet{hetal11}. Both parameters do not affect the
cluster age estimate, which is mainly driven by the difference between the magnitude
at the red clump and that of the Main Sequence turnoff. The position of the red giant branch is
metallicity-dependent, in the sense that the more metal-rich an old star cluster
the redder the red giant branch. Such a behaviour can be blurred by the
age-metallicity degeneracy that mainly affects star clusters younger than
$\sim$ 6 Gyr \citep{os15,piatti2020b}.

We fitted theoretical isochrones  computed by 
\citet[][PARSEC\footnote{http://stev.oapd.inaf.it/cgi-bin/cmd}]{betal12} and 
\citet[][BaSTI\footnote{http://basti-iac.oa-abruzzo.inaf.it/index.html}]{pietrinfernietal2004}.
We used PARSEC isochrones spanning metallicities ([Fe/H]) from -1.5 dex up to -0.5 
dex (Y=0.25), in steps
of 0.2 dex and log(age) from 9.7 up to 9.95 in steps  of 0.25. 
BaSTI isochrones are available only for [$\alpha$/Fe]  = 0.0
dex  (Y=0.25). We fitted these isochrone sets to stars with $P >$ 50 $\%$ allowing shifts  of 
$\Delta$$E(B-V)$= $\pm$ 0.01 mag in order to mitigate zero point offsets between different
isochrone sets, alongside the selective to visual absorption ratios $A_\lambda$/$A_V$ given
by \citet{cetal89} and $A_V/E(B-V)$ = 3.1. As for the age and metallicity uncertainties, we
verified that isochrones differing in $\Delta$(age) = 0.5 Gyr, or $\Delta$[Fe/H] = 0.2 dex, 
result clearly distinguishable when superimposed them on the cluster CMD.
The isochrones which best resemble the cluster CMD features are shown in Fig.~\ref{fig3},
where we also indicated their associated parameter values. From Fig.~\ref{fig3}, we
conclude that KMHK~1592,  located at a mean heliocentric  distance of 49.9 kpc and
affected by a low interstellar reddening ($E(B-V)$=0.042), is a genuine LMC age gap star cluster, with an age 
 ($\sim$ 8.0$\pm$0.5 Gyr) and a metal content ([Fe/H]=-1.0$\pm$0.2 dex)
 similar to those of ESO\,121-SC\,03. 

This second LMC age gap cluster spurs us to speculate on its origin, as well as that of
ESO\,121-SC\,03. Fig.~\ref{fig4} shows the location of both star clusters with
respect to the LMC and LMC cluster populations.
We glimpse three possible scenarios, namely: in-situ formation, stripping
off the SMC, or accretion of a dwarf galaxy. Although the origin of each of these two
star clusters could have in principle been different, their very similar ages and metallicities
favour by chance a common origin. As for the in-situ formation, the LMC star formation history
tells as that the galaxy has experienced a continuous formation of stars out of the available
gas chemically enriched over time. Indeed, \citet{pg13} showed that old and
metal-poor field stars have been preferentially formed in the outer disc, while younger and more 
metal-rich stars have mostly been formed in the inner disc, confirming an outside-in formation. 
Particularly, they provided  evidence for the formation of stars between 5 and 12 Gyr, during the 
star cluster age gap, although chemical enrichment during this period was minimal.
The star formation history maps built by \citet{mazzietal2021} also reveal as a main feature
of the LMC disc a wider and smoother distribution of stellar populations older than $\sim$ 4 Gyr,
which formed at a rate nearly half that of the period of most
intense star formation, which occurred roughly between 4 and 0.5 Gyr ago, at a rate of $\sim$ 
0.3$\msun$/yr \citep[see, also][]{hz09}. Star clusters and field stars 
in the LMC and the SMC have similar age-metallicity relationships \citep{pg13,narlochetal2021,piatti2021f}, so that we would also expect  star clusters 
formed during the age gap, accompanying the field star formation. However,
the existence of only two star clusters of 8-9 Gyr old calls our attention. 
For this reason, we think that an ex-situ origin could explain their appearance in the 
LMC, unless an enough large population of 6-10 Gyr old clusters is discovered in 
the LMC.  Note that the ex-situ origin scenario does not provide any hint to
explain the apparent cease of cluster formation during that period.

The tidal interaction between both Magellanic Clouds can also be a source for star clusters
formed in the SMC were then stripped off by the LMC. Indeed, \citet{carpinteroetal2013} 
modelled the dynamical interaction between both galaxies and their corresponding stellar cluster 
populations, and found that for eccentricities of the orbit of the SMC around the LMC $\ga$ 0.4,
nearly 15 per cent of the SMC star clusters are captured by the LMC, while another 20-50 per cent
is scattered into the intergalactic environment. The star clusters lost by the SMC are the less 
tightly bound ones. These star clusters populate the outer galaxy regions, which have long been
commonly thought to harbour old star clusters. According to the numerical simulations performed
by \citet{carpinteroetal2013} star clusters that originally belonged to the SMC are more likely 
to be found in the outskirts of the LMC. The comparison of the star cluster age-metallicity 
relationships of the LMC and SMC shed light to reconstruct the interaction history between both 
Magellanic Clouds. \citet{p11a} showed that a bursting cluster formation episode took place in the LMC
$\sim$ 2-3 Gyr ago, which has also been detected in the SMC \citet{p11b}. Furthermore, field
star formation in both galaxies have also experienced the chemical enrichment observed in
the star cluster age-metallicity relationships. This means that field star and star cluster
formation were synchronized in the LMC/SMC and that both galaxies interacted at that time.
The SMC exhibits another enhancement of star clusters and field stars at $\sim$ 6-8 Gyr with
a noticeable metallicity spread \citep{p12a}. \citet{tb2009} suggested that such an older
bursting formation episode was caused by the merger with a small gas-rich dwarf. 
Therefore, ESO\,121-SC\,03 and KMHK~1592 could form in the SMC (their ages and metallicities
agree well with the older enhanced formation event) and later captured
by the LMC \citep{carpinteroetal2013}.

More recently, several studies dealt with the LMCformation and
interaction with the SMC, although they do not focus on the cluster formation history. 
To this respect, for 
completeness purposes, we refer the reader to \citet[][; and references therein]{wanetal2020,ruizlaraetal2020,mazzietal2021,williamsetal2021,romanduvaletal2021,gradyetal2021,shippetal2021,cullinaneetal2022}, 
among others.

\citet{mucciarellietal2021} reported that the LMC experienced a merger event in the 
past with a galaxy with a low star formation efficiency and with a stellar mass similar 
to  those of dwarf spheroidal galaxies. Such an LMC satellite completely dissolved into the LMC,
and it was recognized by the peculiar chemical composition of NGC~2005, one of the fifteen
known LMC ancient globular clusters. NGC~2005 would be the only surviving witness, unless
ESO\,121-SC\,03 and KMHK~1592 also had belonged to it. With the aim of confirming this possibility a
chemical tagging of both star clusters is needed, in order to compare their abundances of different
chemical elements with those of LMC field stars for a similar overall metallicity. A detaild
chemistry of both star clusters will definitively address whether the LMC, the SMC or another
dwarf is their progenitor.  

\section{Data availability}

Data used in this work are available upon request to the author.

\section*{Acknowledgements}
 I thank the referee for the thorough reading of the manuscript and
timely suggestions to improve it. 
Based on observations obtained at the international Gemini Observatory, a program of NSF’s NOIRLab, which is managed by the Association of Universities for Research in Astronomy (AURA) under a cooperative agreement with the National Science Foundation. on behalf of the Gemini Observatory partnership: the National Science Foundation (United States), National Research Council (Canada), Agencia Nacional de Investigaci\'{o}n y Desarrollo (Chile), Ministerio de Ciencia, Tecnolog\'{i}a e Innovaci\'{o}n (Argentina), Minist\'{e}rio da Ci\^{e}ncia, Tecnologia, Inova\c{c}\~{o}es e Comunica\c{c}\~{o}es (Brazil), and Korea Astronomy and Space Science Institute (Republic of Korea).










\bsp	
\label{lastpage}
\end{document}